\documentclass[%
 reprint,
 amsmath,amssymb,
 aps,
 prl,
]{revtex4-2}

\usepackage{graphicx}
\usepackage{dcolumn}
\usepackage{bm}
\usepackage{blindtext}
\usepackage{float}
\usepackage{caption,cancel}
\usepackage{cleveref}
\usepackage{subcaption}
\usepackage{xcolor}

\newcommand{\bom}{\boldsymbol{\omega}}

\def \s{\mathbf{s}}
\def \v{\mathbf{v}}
\def \x{\mathbf{x}}

\def \k{\mathbf{k}}
\def \h{\mathbf{h}}

\newcommand*{\CHANGE}[1]{{\color{black}#1}}    
\newcommand*{\GK}[1]{{\color{black}#1}}           

\begin{document}

\preprint{APS/123-QED}

\title{Inverse energy transfer in three-dimensional quantum vortex flows}

\author{P. Z. Stasiak}
\affiliation{School of Mathematics, Statistics and Physics, Newcastle University, Newcastle upon Tyne, NE1 7RU, United Kingdom}

\author{A. Baggaley}
\affiliation{School of Mathematics, Statistics and Physics, Newcastle University, Newcastle upon Tyne, NE1 7RU, United Kingdom}
\affiliation{Department of Mathematics and Statistics, Lancaster University, Lancaster, LA1 4YF, UK}

\author{C.F. Barenghi}
\affiliation{School of Mathematics, Statistics and Physics, Newcastle University, Newcastle upon Tyne, NE1 7RU, United Kingdom}

\author{G. Krstulovic}
\affiliation{Universit\'e C\^ote d'Azur, Observatoire de la C\^ote d'Azur, CNRS,Laboratoire Lagrangre, Boulevard de l'Observatoire CS 34229 - F 06304 NICE Cedex 4, France}

\author{L. Galantucci}
\affiliation{Istituto per le Applicazioni del Calcolo ``M. Picone" IAC CNR, Via dei Taurini 19, 00185 Roma, Italy}

\date{\today}

\begin{abstract}
Vortex reconnections play a fundamental role in fluids.
They increase the complexity of flow and develop small-scale motions.
In this work, we report that in superfluids, they can also excite large scales.
We numerically illustrate that during a superfluid vortex reconnection energy 
is injected into the thermal (normal) component of helium~II at small length scales, but is transferred nonlinearly  to larger length scales, increasing the integral length scale of the normal fluid. 
We show, \GK{by studying about fifty different reconnections,} that this inverse energy transfer is triggered by the helical imbalance generated 
in the normal fluid flow by
the mutual friction force coupling the superfluid vortices and the normal component. 
We finally discuss
the relevance of our findings to 
the problem of superfluid turbulence.
\end{abstract}

\maketitle

\begin{figure*}
	\centering
  \includegraphics[width=0.95 \textwidth]{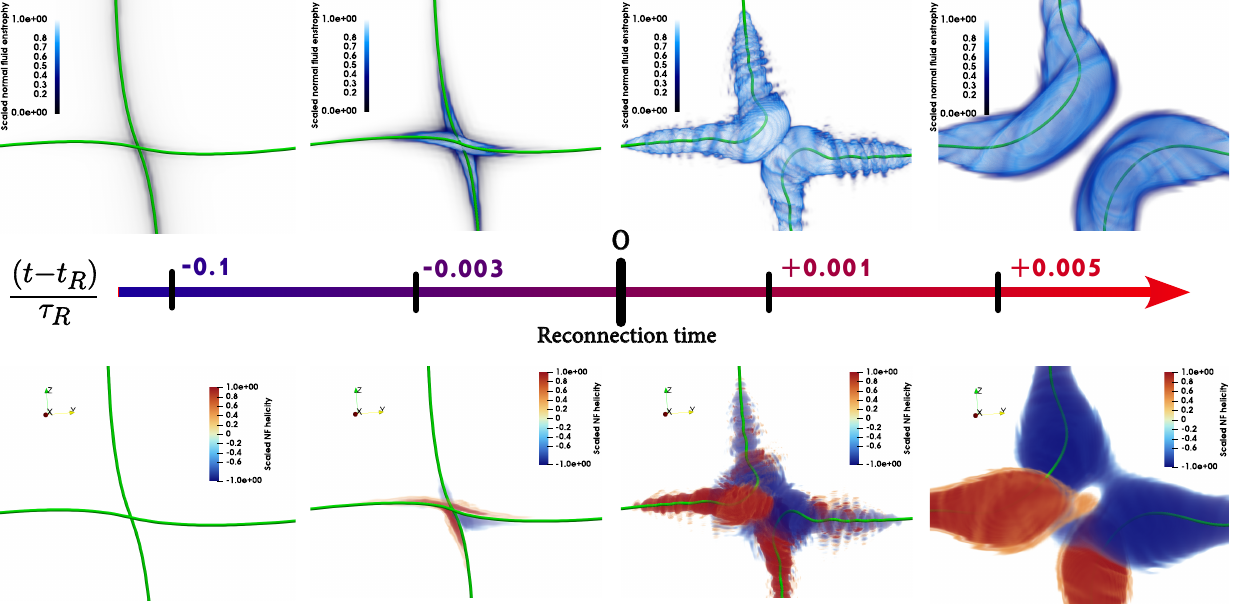}

	\caption{
\CHANGE{
Three-dimensional rendering of the time evolution of
an initially orthogonal superfluid vortex configuration 
undergoing a reconnection at $t=t_R$. Temperature is $T = 1.9$K and time is made
dimensionless with 
$\tau_R = L^2/\kappa$. }
Green tubes represent the superfluid vortex lines (the tubes’ radii have been 
greatly exaggerated for visual purpose). In the top sequence,
the blue volume rendering represents the scaled normal fluid enstrophy 
$\bom^2/\bom^2_{max}$. Note the Kelvin wave on the superfluid
vortex at $(t-t_R)/\tau_R\approx0.001$. In the bottom sequence, the red/blue volume
rendering at the same times represent scaled positive/negative normal fluid
helicity.	
}
\label{fig:visualisation}
\end{figure*}

Turbulence is ubiquitous in the universe.  It occurs in systems
as large as nebulae of interstellar gas, and as small as clouds of
few thousands atoms confined by lasers in the laboratory.
Turbulence shapes patterns and properties of fluids of all
kinds, from ordinary
 viscous fluids (Navier-Stokes turbulence\cite{frisch1995}) 
to electrically conducting fluids (magneto-hydrodynamics turbulence
\cite{canuto-dalsgaard-1998}) to quantum fluids
(quantum turbulence \cite{barenghi-etal-2023,
Barenghi_Skrbek_Sreenivasan_2023}).
All turbulent systems are characterised by 
the existence of a wide range 
of length scales across which inviscid conserved quantities 
are transferred without loss in the spirit of the cascade 
depicted by Richardson \cite{richardson1922weather}. 

In three-dimensional (3D) classical fluids, turbulence 
is characterised by a direct (forward) cascade: the
non-linear dissipationless transfer of kinetic energy from the scale of
the large eddies (at which energy is injected) to the smallest length scales
at which energy is dissipated into heat
\cite{richardson1922weather,kolmogorov-1941}. 
The resulting distribution of energy across length scales is
the celebrated Kolmogorov energy spectrum 
\cite{kolmogorov-1941,frisch1995}. 

Confining Navier-Stokes turbulence to two-dimensions (2D) entails 
fundamentally distinct physics: a dual cascade emerges of energy and enstrophy 
(mean squared vorticity) \cite{kraichnan-1967,boffetta-ecke-2012}, 
the two conserved quantities in ideal two-dimensional flows.
While the enstrophy cascade is direct (from large to small
scales), the energy cascade is inverse (from small to large scales)
\cite{boffetta-musacchio-2010}. This inverse cascade 
may favour the generation and persistence of large coherent 
structures \cite{laurie-etal-2014}. 

Remarkably, the same cascade phenomenology is observed in turbulent flows 
of quantum fluids, {\textit{i.e.} fluids at very low temperatures whose physics is
dominated by quantum effects.
Examples of such fluids are superfluid helium 
and atomic Bose-Einstein Condensates (BECs). 
The dynamics of these systems can be successfully depicted in terms of 
a two-fluid model \cite{tisza-1938,landau-1949,skrbek-sreenivasan-2012} 
describing the quantum fluid as the mixture of two components, 
the superfluid component and the thermal (or normal) component, which 
interact by means of a mutual friction force 
\cite{jackson-etal-2009,hall-vinen-1956a,hall-vinen-1956b}. 
The superfluid component flows without viscosity and
vanishing entropy; its vorticity is
confined to effectively one-dimensional vortex filaments
of atomic core thickness (called quantum vortices or vortex lines), 
around which the circulation of the velocity is quantised.
In BECs the thermal component forms a ballistic gas,
whereas in superfluid $^4$He it can be described as
a classical viscous fluid.
Despite these significant differences with respect to ordinary fluids, 
the direct
kinetic energy cascade has
indeed been observed in three-dimensional superfluid
turbulence 
\cite{maurer1998,salort2010turbulent,baggaley2012,sherwin-robson2015,Muller_KolmogorovKelvinWave_2020,Muller_IntermittencyVelocityCirculation_2021}.
Evidence of this 
direct
cascade has been found also in
three-dimensional turbulent BECs \cite{middleton-spencer2022}. 
In this forward energy cascade a fundamental role is played by the dynamics and interactions of quantum vortices,
in particular by their reconnections, phenomena where  
two vortex lines collide and recombine, exchanging heads and tails, 
altering the overall topology of the flow
\cite{koplik-levine-1993,bewley-etal-2008,rorai-etal-2016,serafini-etal-2017,galantucci-baggaley-parker-barenghi-2019,villoisUniversalNonuniversalAspects2017,villois2020irreversible}.
Reconnections occur frequently in 3D quantum turbulence and are essential to
the energy transfer towards small scales by engendering the breakdown of coherent vortex structures
and triggering the Kelvin-wave cascade \cite{vinen-2001}.
In 2D, similarly to classical turbulence, 
an inverse energy cascade characterises turbulence in two-dimensional BECs, 
as shown in theoretical \cite{bradley2012energy,reeves2013,simula2014emergence,Muller_ExploringEquivalenceTwoDimensional_2024} and experimental \cite{johnstone2019evolution,gauthier2019giant} studies.

In turbulent systems, the type and the number of sign-defined ideal invariants determine the direction of cascades. Indeed, the famous 
Fjørtoft argument \cite{fjortoft1953changes} predicts 
the existence of an inverse energy cascade
in 2D classical turbulence. 
It also predicts an inverse particle and a direct energy cascade 
for 3D wave turbulent BECs, as recently addressed theoretically \cite{Zhu_DirectInverseCascades_2023}. In 3D classical fluids, helicity, which is also an inviscid invariant, is not sign-defined and thus only a direct energy cascade is possible. However, recent studies have demonstrated that the direction of the energy 
cascade may be inverted by artificially controlling the chirality of the 
flow, \textit{i.e.} the balance between positive and negative helical 
modes \cite{moffatt1969}.
Indeed, by restricting the non-linear energy transfer to homochiral 
interactions via a suitable decimation of the Navier-Stokes equation 
\cite{biferaleInverseEnergyCascade2012a,biferale-etal-2013}, by
controlling the weight of homochiral interactions \cite{sahoo-etal-2017},
or by the external injection 
of positive helical modes at all length scales 
\cite{plunianInverseCascadeEnergy2020a}, inverse energy cascades 
have been observed in three-dimensional turbulence of classical fluids. 
In brief, when the flow is synthetically designed to have an 
enhanced chirality, an inverse energy cascade can observed.

In this work, we unveil a similar dynamics occurring in superfluid helium
($^4$He) as a result of vortex reconnections. 
We show that the mutual friction force arising from the  
vortex reconnection is chiral, injecting in the normal fluid prevalently 
helicity of a given sign 
Thus, as a consequence of vortex reconnections,
we observe an increase of the chiral imbalance of the quantum fluid, producing a transfer of kinetic energy from small to large scales, similarly to the phenomenology observed in 3D helically-decimated classical flows. 
 Unlike classical fluids, such a chiral imbalance arises naturally as a physical process in the normal fluid.

%



To model superfluid helium dynamics, we employ the recently developed FOUCAULT model
\cite{galantucciNewSelfconsistentApproach2020b}.  In this approach, superfluid vortex lines are
parametrized as one-dimensional space curves  $\s(\xi,t)$, $\xi$ and $t$ being arclength and time respectively, exploiting the large separation of length scales between the vortex core radius, the Lagrangian discretisation along the vortex lines $\Delta\xi$, and the average radius of curvature $R_c$ of the vortex lines. The vortex lines evolve according to the following equation of motion:
\begin{equation}
    \dot{\s}(\xi,t) = \v_s + \frac{\rho_n}{\rho}\left[\v_{ns}\cdot\s'\right]\s' + \beta\s'\times\v_{ns} + \beta'\s'\times\left[\s'\times\v_{ns}\right],
\end{equation}
where $\dot{\s} = \partial\s/\partial t$, $\s' = \partial\s/\partial\xi$ is the unit tangent vector, $\v_n$ and $\v_s$ are the normal fluid and superfluid velocities at $\s$, $\v_{ns} = \v_n-\v_s$, and $\beta,\, \beta'$ are temperature and Reynolds number dependent mutual friction coefficients \cite{galantucciNewSelfconsistentApproach2020b}. The calculation of the superfluid velocity $\v_s$ is performed via the computation of the Biot-Savart integral de-singularised with standard techniques (see Supplementary Material \cite{suppMat}). The normal fluid is described classically using the incompressible ($\nabla\cdot\v_n=0$) 
Navier-Stokes equation
\begin{equation}
    \frac{\partial\v_n}{\partial t} + (\v_n\cdot\nabla)\v_n = -\frac{1}{\rho}\nabla p  + \nu_n\nabla^2\v_n + \frac{\mathbf{F}_{ns}}{\rho_n} \; \; , 
\end{equation}
where $\rho_n$ and $\rho_s$ are the normal fluid and superfluid densities,
$\rho=\rho_n + \rho_s$,  $p$ is the pressure,  $\nu_n$ is the kinematic 
viscosity of the normal fluid, and the mutual friction force per unit
volume, $\mathbf{F}_{ns}$, is the line integral of the mutual friction 
force per unit length, $\mathbf{f}_{ns}$ \cite{suppMat}:
\begin{equation}
\mathbf{F}_{ns}(\x) = 
\oint_{\mathcal{C}}\delta(\x-\s)\mathbf{f}_{ns}(\s)d\xi,     
\end{equation}
$\mathcal{C}$ representing the entire vortex configuration. 
The regularisation of mutual friction is performed using a physically self-consistent scheme \cite{galantucciNewSelfconsistentApproach2020b}. 
We consider a periodical box of size $L=2\pi$ (so that  wavevectors are integers).

To study the reconnection dynamics, we consider 
\CHANGE{
two distinct initial conditions at temperatures $T=1.9K$ and $T=2.1K$: (a) a pair of initially orthogonal vortices (a second pair of orthogonal vortices, with
corresponding vortices of each pair having opposite circulation, is included in the computation in order to preserve periodicity along the boundaries \cite{suppMat};  
the vortex pairs are separated by the distance $D_{\ell}$, while each vortex within each pair is initially at distance $d_{\ell}$ to the other vortex, such that $d_{\ell}\ll D_{\ell}$ in order to ensure that the dynamics in the vicinity of the reconnection is dominated by local interactions); (b) a set of 49 Hopf-links, 
each consisting in two, perpendicular, linked vortex rings at varying initial distances from each other \cite{suppMat}.
}\GK{Note that the two types of initial conditions are topologically very different, as for the Hopf-Link the superfluid initial helicity is non-zero due to their linking (it is zero for the orthogonal case). For both configurations, the initial normal fluid helicity is zero.}


The evolution of the reconnection of an initially orthogonal single vortex pair is reported in Fig.~\ref{fig:visualisation}. 
The first row shows the reconnecting superfluid vortices (in green) accompanied by normal fluid structures generated by the mutual friction, 
here displayed as enstrophy rendering $\bom(\x)^2=|\nabla\times \v_n|^2$. Such structures are the signature of the violent irreversible energy transfers in vortex reconnections 
\cite{stasiak2025experimental}. The second row of Fig.~\ref{fig:visualisation} shows the rendering of the local helicity density $H(\x)=\v_n\cdot\bom$: we observe a clear local helicity production, with an abrupt change of sign due to the rearrangement of the vortex topology. Remarkably, during the reconnection process there is a sudden net overall
normal fluid helicity production, as shown in Fig.~\ref{fig:total-helicity} (a). 
\CHANGE{
Indeed, the temporal evolution of the total normal fluid helicity 
$\mathcal{H} = \int_{\mathcal{V}}H(\x)dV$ computed over the entire volume $\mathcal{V}$ shows that at reconnection time $t_R$ there is a significant overall 
injection of helicity of a given sign. In particular, in the reconnections we monitor the injected helicity is negative (positive) for the orthogonal (Hopf-link) 
reconnections.
}
\begin{figure}[t]
    \centering

    \includegraphics[width=0.49\textwidth]{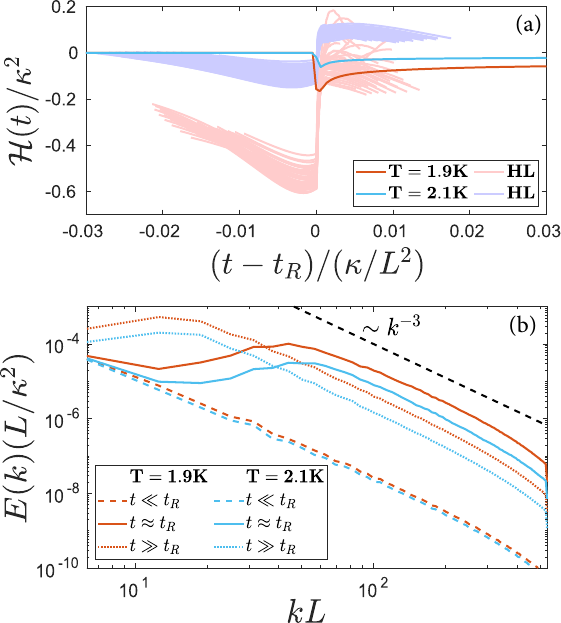}
        
\caption{\CHANGE{ \emph(a): Temporal evolution of the total normal fluid helicity $\mathcal{H}$, normalised by the quantum of circulation $\kappa$. 
Bold solid lines refer to orthogonal reconnections, slightly faded lines refer to the set of Hopf-links reconnections. Red color refers to $T=1.9K$, blue to $T=2.1K$.
\emph(b): Normal fluid kinetic energy spectrum $E(k)$ before 
reconnection (dashed lines), at reconnection (solid lines) 
and after reconnection (dotted lines), for a pair of initially orthogonal vortices. Red (blue) lins correspond to  $T=1.9K$ ($T=2.1K$).}
}
\label{fig:total-helicity}
\end{figure}

We now focus on the time evolution of the normal fluid energy spectrum $E(k)$, defined by 
\begin{equation}
    E = \frac{1}{(2\pi)^3}\int_{\mathcal{V}}\frac{1}{2}|\v_n|^2 dV = \int_0^{\infty}E(k)dk
\end{equation}
where $E$ is the total normal fluid energy and $k$ is the magnitude of the three-dimensional wavenumber. 
The energy spectrum $E(k)$ for orthogonal reconnections is displayed in Fig.~\ref{fig:total-helicity} (b) (corresponding plots for the
Hopf-links reconnections show a similar behaviour \cite{suppMat}).
It clearly emerges that, during the reconnection, energy is predominantly injected into the normal fluid at intermediate and small length scales. 
For $kL>30$ in correspondence of the reconnection time $t_R$,
we observe a significant increase of the normal fluid energy spectral density:
$E(k,\, t\approx t_R)/E(k,\, t\ll t_R) \approx 10^2$. 
In the post-reconnection regime, we simultaneously observe a small 
decrease of the spectrum at intermediate 
and small scales ($kL > 30$) and an increase at large scales, 
suggesting the existence of a mechanism by 
which energy generated at small length scales is transferred to larger scales. 
 
\begin{figure}
    \centering
    \includegraphics[width=0.49\textwidth]{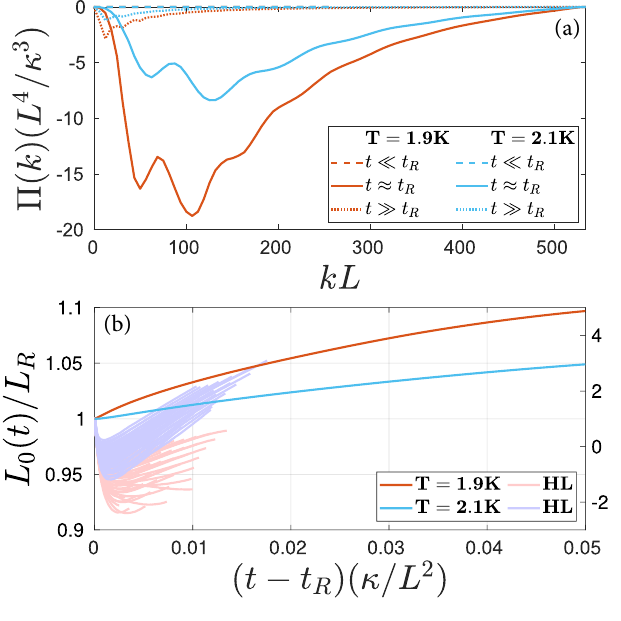}


    \caption{\CHANGE{
    \emph{(a)}: Spectral normal fluid kinetic energy flux, $\Pi (k)$ for orthogonal reconnections. 
    Times and temperatures are labelled as in Fig.~\ref{fig:total-helicity} (b).
    \emph{(b)}: Post reconnection temporal evolution of the integral length scale, $L_0(t)$ scaled by the reconnection time integral scale $L_R=L_0(t_R)$. 
    Slightly faded lines refer to the set of Hopf-links reconnections (left axis) and bold solid lines refer to orthogonal reconnections (right axis). 
    Red (blue) lins correspond to  $T=1.9K$ ($T=2.1K$).}
    }
\label{fig:spectra}
\end{figure}

To shed light on this mechanism, as customary for turbulent flows, we analyse the spectral energy flux 
\begin{equation}
    \Pi(k)=\int_{|{\bf p}|<k}  \hat{\mathbf{v}}_n^*\cdot \left [\widehat{(\v_n\cdot\nabla)\v_n} \right ]d {\bf p} +c.c. \;\; ,
\end{equation}
where $\widehat{\cdot}$ indicates the Fourier transform. The computation of $\Pi(k)$ is reported in Fig.~\ref{fig:spectra} (a) 
for orthogonal reconnections (similar behaviour is observed in Hopf-links reconnections \cite{suppMat}).

We observe that $\Pi(k) < 0$ for all $k$ during and after reconnection;
we also observe that, near the time of reconnection,
the peak value of $|\Pi(k)|$ is in the range \CHANGE{$50 < kL <150$}.
The negative sign of $\Pi(k)$ is evidence of a flux of kinetic  energy from small to large scales, exciting larger and larger scales. 
This behaviour is quantified by the temporal evolution of the integral length scale $L_0(t)$, defined as
\begin{equation}
    L_0(t) = \frac{\pi}{2 E}\int_0^{\infty}\frac{E(k,t)}{k}dk, 
\end{equation}
\CHANGE{
and reported in Fig.~\ref{fig:spectra} (b). The figure clearly shows that
$L_0$ indeed increases steadily in the post-reconnection regime after a small transient in the Hopf-links configuration.}

To explain the inverse energy transfer shown in Fig.~\ref{fig:spectra} (b),
we look whether the reconnection triggers a chirality imbalance in the flow.  
We decompose the incompressible Fourier modes of the normal fluid velocity 
into helical modes \cite{waleffe-1992}:
\begin{equation}
\hat{\mathbf{v}}_n (\k) = \hat{\mathbf{v}}_n^+(\k) +\hat{\mathbf{v}}_n^-(\k)=
 v_n^+(\mathbf{k}) \mathbf{h}^+(\mathbf{k})+v_n^-(\mathbf{k}) \mathbf{h}^-(\mathbf{k}),
\end{equation}
where $\mathbf{h}^\pm (\mathbf{k})$ are the two eigenvectors of the curl 
operator, \textit{i.e.} $i\k~\times~\h^{\pm}(\k)~=~\pm k \h^{\pm}(\k)$. 
Similarly, we decompose the transverse modes of the mutual friction force:
$\hat{\mathbf{F}}_{ns}^{\perp}(\k) = f^+(\k) \mathbf{h}^+ + f^-(\k) \mathbf{h}^-$
(the  Fourier modes of $\mathbf{F}_{ns}$ parallel to the wavemumber 
$\k$ do not play any role in the time evolution of $\mathbf{v}_n$ 
due to the incompressible constraint). Finally, the helical decomposition naturally allow us decompose the total helicity as $\mathcal{H}=\mathcal{H}^+-\mathcal{H}^-$ \cite{plunianInverseCascadeEnergy2020a}.
A chiral imbalance occurs if the mutual friction force is helical, 
\textit{i.e.} if $\mathcal{F}_R \neq 1$, where
\begin{equation}
	\mathcal{F}_R(t) = \frac{\int dk |f^+(k,t)|^2}{\int dk |f^-(k,t)|^2},
\end{equation}
with $|f^\pm|^2$ the squared norm of the mutual friction
helical decomposition coefficients. 
In Fig.~\ref{fig:mutual-friction-decomp}, we show the temporal evolution of $\mathcal{F}_R$ for both temperatures, for orthogonal reconnections
(Hopf-links reconnections show a similar behaviour \cite{suppMat}). 
\begin{figure}[h!]
    \centering
    \includegraphics*[width=0.48\textwidth]{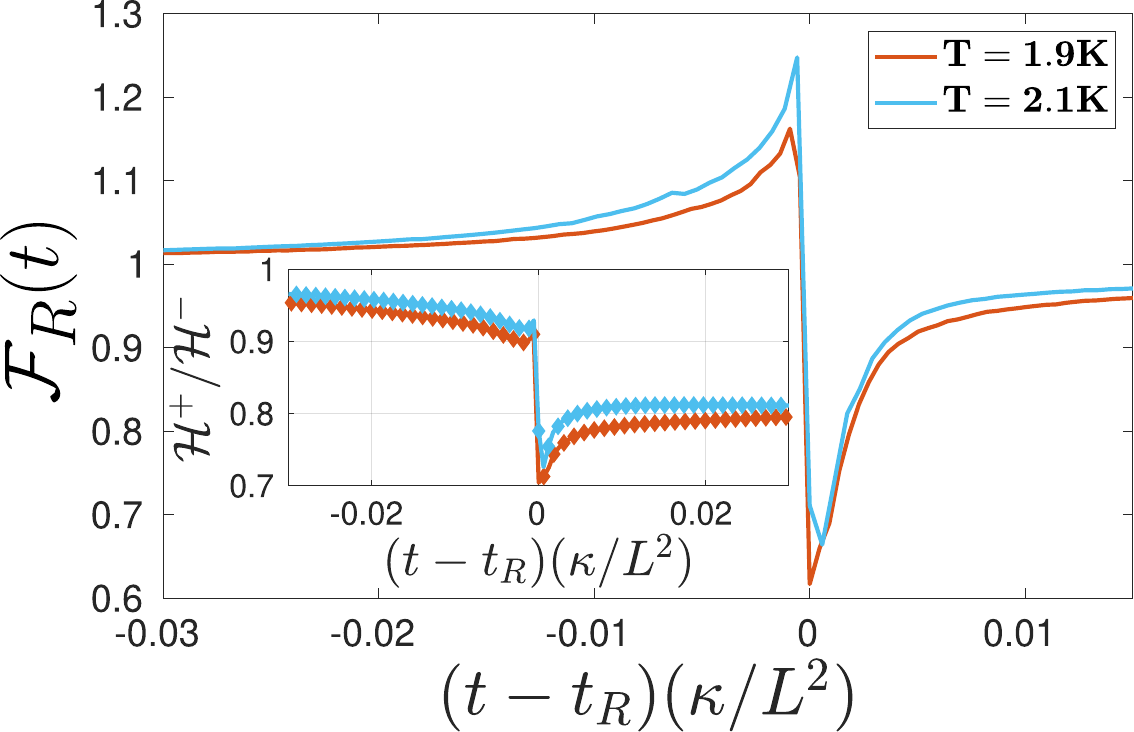}
    \caption{Temporal evolution of the ratio of helically-projected mutual friction force components $f^\pm$, for orthogonal reconnections. 
    Inset: temporal evolution of total helical components.}
    \label{fig:mutual-friction-decomp}
\end{figure}
It is apparent
that during and after the reconnection, the mutual friction force is strongly chiral,
injecting more negative helicity than positive helicity. 
As a result, the ratio $\mathcal{H}^+/\mathcal{H}^-$ 
(reported in the inset of Fig.~\ref{fig:mutual-friction-decomp}) 
decreases significantly at reconnection and remains smaller than unity even at later times, 
indicating that the flow is persistently chiral. We conclude that the reconnection triggers indeed a chiral imbalance, \CHANGE{leading to the observed inverse energy transfer}.
For the sake of completeness, it is worth noting that the total helicity of the flow (obtained by adding the superfluid helicity to that of the normal fluid $\mathcal{H}$) is not conserved.
From Fig.~\ref{fig:mutual-friction-decomp}, we consider a time $t^*$ after reconnection where the ratio $\mathcal{F}_R$ has sufficiently decayed to a roughly 5\% difference and we determine the non-dimensional timescale $\tau = (t^*-t_R)/\tau_R$ during which the mutual friction force is chiral as a result of reconnections: $\tau \approx 0.01$ and $\tau \approx 0.005$
for $T=1.9$K and $T=2.1$K, respectively, corresponding dimensionally to $\tau \approx 0.1$s for both temperatures. 
In superfluid turbulence, the timescale between two consecutive reconnections can be smaller than $\tau$ provided that the
vortex line density $\mathcal{L}$ (length of vortices per unit volume) is larger than $10^8 \text{m}^{-2}$ \cite{stasiak2025experimental,barenghi2004},
a condition which is easily met in superfluid helium experiments \cite{roche2007,Babuin2014}.

In conclusion, the reconnection of quantum vortices in the two-fluid regime
($T\gtrsim 1.5$K) not only injects punctuated
energy in the normal fluid \cite{stasiak2025experimental}, but also triggers in the normal fluid
a transfer of kinetic energy towards the large scales. 
This inverse energy transfer is triggered by the helical character of the
mutual friction, directly arising from the Kelvin waves released by the reconnecting
cusp, which produces a chiral imbalance in the normal fluid, as 
previously observed in turbulent Navier-Stokes flows
\cite{biferaleInverseEnergyCascade2012a,plunianInverseCascadeEnergy2020a}.
\CHANGE{
This inverse energy transfer contrasts with the direct energy transfer observed 
in classical vortex reconnections \cite{yao-hussain-2020} showing that vortex reconnections 
in quantum fluids may show similarities and differences with their classical counterpart 
depending on the fluid component investigated: if we consider the approach and separation 
of superfluid vortex filaments, we recover a classical behaviour
\cite{stasiak2025experimental}, while if we analyse the normal fluid energy transfer we
observe a strong non-classical effect. \GK{The robusteness of our results has been demonstrated by studying almost fifty different reconnections having an initial different topology.} The results presented hence contribute to 
identifying similarities and differences between classical and quantum turbulence 
\cite{galantucci-2025}. 
}

Our findings have profound implications for the nature of turbulence in 
finite temperature superfluids, where vortex reconnections play a key role. 
In circumstances where the vortex density 
$\mathcal{L}$ is large and where the isotropy of the vortex tangle 
is broken by external forcing (for instance thermal counterflows~\cite{vinen-1957a}), 
the chirality of the flow generated by the 
frequent, \CHANGE{likely} non-symmetrical, reconnections may be strong enough to induce an inverse energy cascade~\cite{notes_recon}.
The microscopic mechanism that we have described may be relevant in triggering
the inverse energy cascade which is observed numerically in  
large-scale simulations of counterflow turbulence at large heat fluxes \cite{polanco2020}, which is indeed inherently not isotropic. However, formalising the connection between high counterflow, unbalanced helicity injection and inverse energy cascade is  challenging, as the model used in \cite{polanco2020} and the one used in our work describe superfluid helium at very different scales.
Our work hence motivates further detailed studies of
the role played by helicity in superfluid dynamics 
\cite{dileoni2016,galantucci2021},
moving the emphasis from few vortex systems \cite{scheeler2014} 
to fully coupled superfluid turbulence.

\begin{acknowledgments}
    G.K. was supported by the Agence Nationale de la Recherche through the project the project QuantumVIW ANR-23-CE30-0024-02.
    This work has been also supported by the French government, through the UCAJEDI Investments in the Future project managed by the National Research Agency (ANR) with the reference number ANR-15-IDEX-01. P.Z.S. acknowledges the financial support of the UniCA ``visiting doctoral student program'' on complex systems. Computations were carried out at the Mésocentre SIGAMM hosted at the Observatoire de la Côte d’Azur.
\end{acknowledgments}


\begin{thebibliography}{59}
\bibitem [{not()}]{notes_recon}%
  \BibitemOpen
  \href@noop {} {\bibinfo {title} {The symmetry breaking of the vortex
  configuration is a key element, as perfectly symmetrical reconnections (with
  opposite vortex line orientations) would inject net helicities of opposite
  sign leading globally to a non-chiral flow.}}\BibitemShut {Stop}%
\end{thebibliography}

%

\appendix 

\section*{Supplementary Materials}

\subsection{Numerical Methods}

\begin{figure}
	\centering
	\begin{subfigure}[b]{0.49\textwidth}
		\centering
		\includegraphics[width=0.8\textwidth]{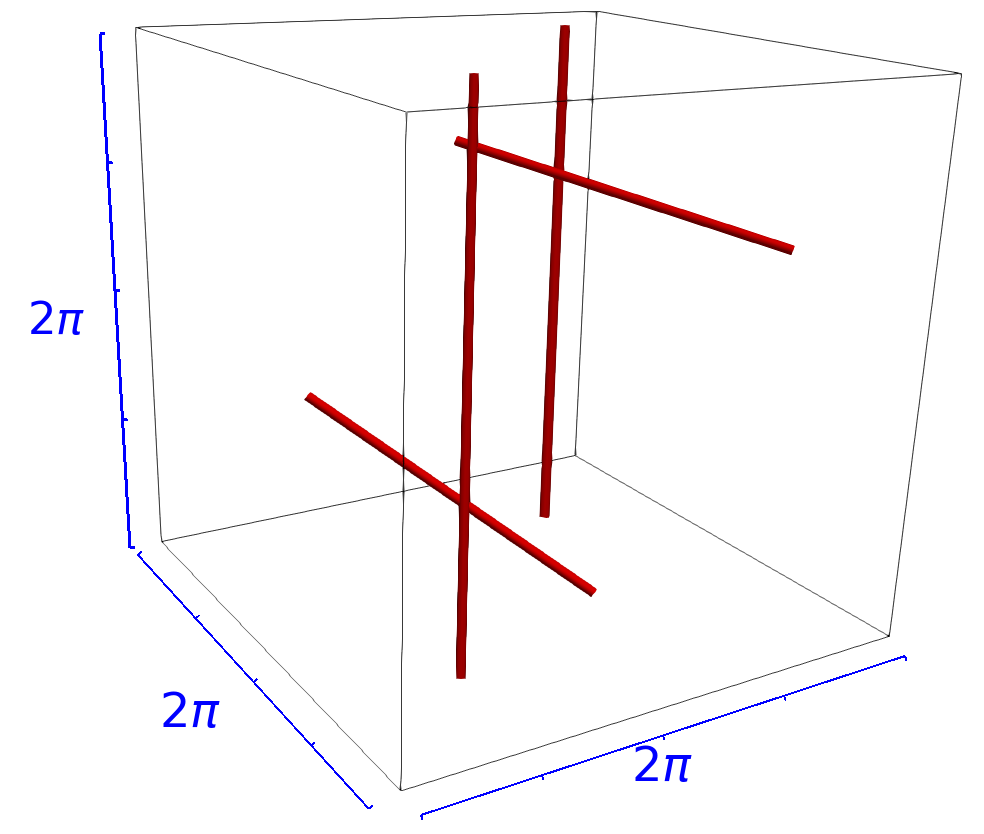}
	\end{subfigure}
	\begin{subfigure}[b]{0.49\textwidth}
		\centering
		\includegraphics[width=0.9\textwidth]{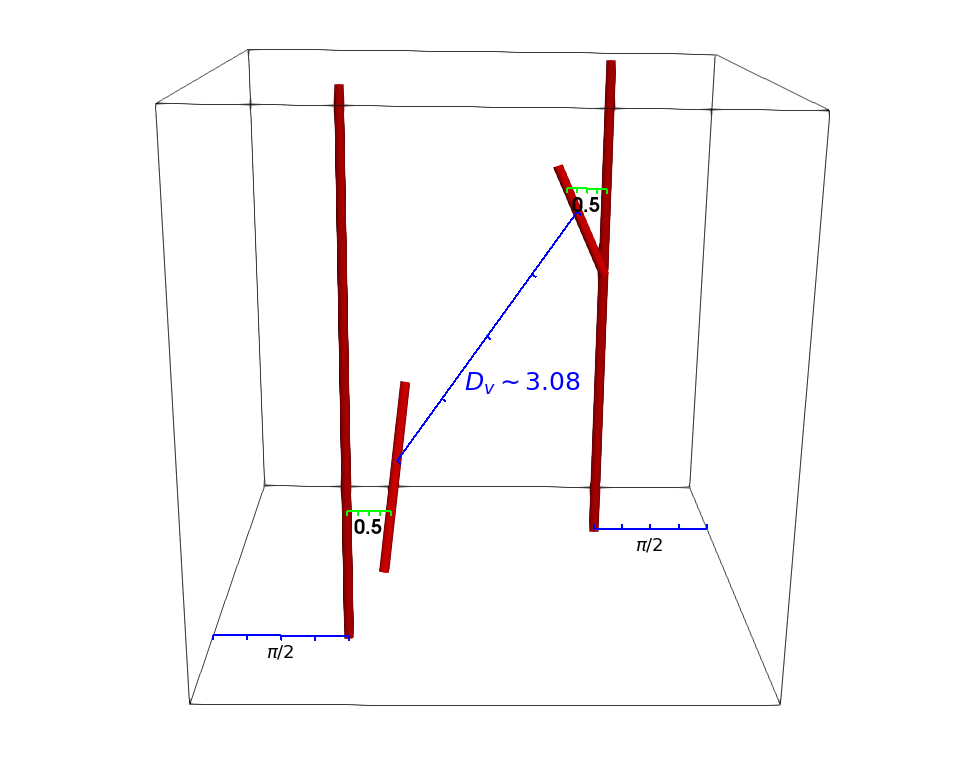}
	\end{subfigure}
	\caption{Schematic diagram of the orthogonal vortex configuration.}
	\label{fig:schematic}
\end{figure}

Using Schwarz mesoscopic model \cite{schwarzThreedimensionalVortexDynamics1988a}, vortex lines can be described as space curves $\s(\xi,t)$ of infinitesimal thickness, with a single quantum of circulation $\kappa=h/m_4=9.97\times10^{-8}\text{m}^2/\text{s}$, where $h$ is Planck's constant, $m_4=6.65\times10^{-27}\text{kg}$ is the mass of one helium atom, $\xi$ is the natural parameterization, arclength, and $t$ is time. These conditions are a good approximation, since the vortex core radius of superfluid \textsuperscript{4}He($a_0=10^{-10}\text{m}$) is much smaller than any of the length scales of interest in turbulent flows. The equation of motion is
\begin{equation}
	\dot{\s}(\xi,t) = \v_s + \frac{\rho_n}{\rho}\left[\v_{ns}\cdot \s'\right]\s' + \beta\s'\times\v_{ns}+\beta'\s'\times\left[\s'\times \v_{ns}\right],
\end{equation}
where $\dot{\s}=\partial\s/\partial t$, $\s'=\partial\s/\partial \xi$ is the unit tangent vector, $\v_{ns}=\v_n - \v_s$, $\v_n$ and $\v_s$ are the normal fluid and superfluid velocities at $\s$ and $\beta$,$\beta'$ are temperature and Reynolds number dependent mutual fricition coefficients \cite{galantucciNewSelfconsistentApproach2020b}. The superfluid velocity $\v_s$ at a point $\x$ is determined by the Biot-Savart law
\begin{equation}
	\v_s(\x,t) = \frac{\kappa}{4\pi}\oint_{\mathcal{T}}\frac{\s'(\xi,t)\times\left[\x-\s(\xi,t)\right]}{|\x-\s(\xi,t)|}d\xi,
\end{equation}
where $\mathcal{T}$ represents the entire vortex configuration.
There is currently a lack of a well-defined theory of vortex reconnections in superfluid helium, like for the Gross-Pitaevskii equation \cite{villoisIrreversibleDynamicsVortex2020,villoisUniversalNonuniversalAspects2017a,promentMatchingTheoryCharacterize2020a}. An \emph{ad hoc} vortex reconnection algorithm is employed to resolve the collisions of vortex lines \cite{baggaleySensitivityVortexFilament2012a}.

A \emph{two-way model} is crucial to understand the accurately interpret the back-reaction effect of the normal fluid on the vortex line and vice-versa \cite{stasiakCrossComponentEnergyTransfer2024}. We self-consistently evolve the normal fluid $\v_n$ with a modified Navier-Stokes equation
\begin{equation}
	\frac{\partial \v_n}{\partial t} + (\v_n\cdot\nabla)\v_n = -\nabla\frac{p}{\rho} + \nu_n\nabla^2\v_n + \frac{\mathbf{F}_{ns}}{\rho_n},
\end{equation}
\begin{equation}
	\mathbf{F}_{ns} = \oint_{\mathcal{T}}\mathbf{f}_{ns}\delta(\x-\x)d\xi, \quad \nabla\cdot\v_n=0,
\end{equation}
where $\rho=\rho_n + \rho_s$ is the total density, $\rho_n$ and $\rho_s$ are the normal fluid and superfluid densities, $p$ is the pressure, $\nu_n$ is the kinematic viscosity of the normal fluid and $\mathbf{f}_{ns}$ is the local friction per unit length \cite{galantucciCoupledNormalFluid2015a}
\begin{equation}
	\mathbf{f}_{ns} = -\mathcal{D}\s'\times\left[\s'\times(\dot{\s}-\v_n)\right]-\rho_n\kappa\s'\times(\v_n-\dot{\s}), 
	\label{eq:mutual-friction}
\end{equation}

\begin{figure*}
	\centering
	\includegraphics[width=0.7\textwidth]{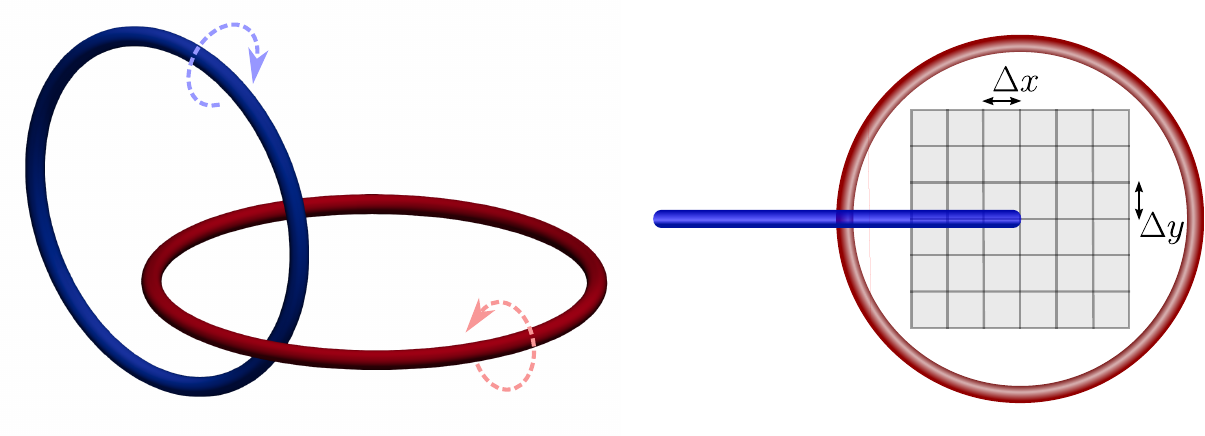}
	\caption{Schematic diagram of the Hopf link vortex configuration.}
	\label{fig:Hopf-link}
\end{figure*}

where $\mathcal{D}$ is a coefficient dependent on the vortex Reynolds number and intrinsic properties of the normal fluid. The regularization of the mutual friction force onto the normal fluid grid is physically motivated by the strongly localized injection of vorticity during the momentum exchange of point-like particles and viscous flow in classical fluid dynamics \cite{gualtieri2015exact,gualtieri2017turbulence}. In short, the localized vorticity induced by the relative motion between the vortex lines and the normal fluid is diffused to discretization of the grid spacing $\Delta x$ in a time interval $\epsilon_R$. In this way, the delta-forced friction as defined in Eq.~\ref{eq:mutual-friction} is regularized by a Gaussian function, the fundamental solution of the diffusion equation. Further details of the method for classical fluids are contained in \cite{gualtieri2015exact,gualtieri2017turbulence} and for FOUCAULT in \cite{galantucciNewSelfconsistentApproach2020b}. 

In this Letter, we report all results using dimensionless units, where the characteristic length scale is $\tilde{\lambda} = D/D_0$, where $D^3=(1\times10^{-3}\mathrm{m})^3$ is the dimensional cube size, $D_0^3=(2\pi)^3$ is the non-dimensional cubic computational domain. The time scale is given by $\tilde{\tau}=\tilde{\lambda}^2\nu_n^0/\nu_n$, where the non-dimensional viscosity $\nu_n^0$ resolves the small scales of the normal fluid. In this work, we consider two vortex configurations - initially orthogonal vortices and Hopf links.\\

\paragraph{Orthogonal reconnection:} The characteristic quantities are $\tilde{\lambda}=1.59\times10^{-4}$m, $\nu_n^0=0.32$ and $\tilde{\tau}=0.366$s at $T=1.9K$ and $\tilde{\tau}=0.485$s at $T=2.1K$. The vortices are initialized as two pairs of orthogonal vortices, as shown in the schematic of Fig.~\ref{fig:schematic}. The separation between vortices in each pair $d$ is set to be $d_v=0.5$ in dimensionless units, and the shortest distance between pairs is $D_v=\sqrt{(\pi-d_v/2)^2+\pi^2}\approx 3.08$, so that $d_v\ll D_v$. The Lagrangian discretization of vortex lines is $\Delta \xi = 0.025$ (a total of 1340 discretization points across the 4 vortex lines), using a timestep of $\Delta t_{VF} = 5.56\times10^{-6}$. For the normal fluid, a total of $N=256^3$ mesh point were used, with a timestep of $\Delta t_{NS} = 45\Delta t_{VF}$. \\

\paragraph{Hopf link:} The characterstic quantities are $\tilde{\lambda}=1.59\times10^{-4}$m, $\nu_n^0=0.16$ and $\tilde{\tau}=0.1836$s at $T=1.9K$ and $\tilde{\tau}=0.2439$s at $T=2.1K$. Vortices are initialized as shown in Fig.~\ref{fig:Hopf-link}, where the blue vortex ring is chosen at an initial offset $n_x\Delta x$ and $n_y \Delta y$ where $n_x,n_y\in\lbrace-3,\cdots 3\rbrace$ and $\Delta x = \Delta y = 0.125$ in units of the code. This gives a total of 49 individual reconnections for each temperature. Both of the rings have radius $R\approx 1$ also in units of the code. Furthermore, each reconnection is supplemented with a normal fluid ring around the superfluid vortex ring, which is generated by superimposing a normal fluid ring generated by a propagating ring of the same radius. In this way, we eliminate a transient phase of generating normal fluid structures. The Lagrangian discretization of the vortex lines is $\Delta\xi = 0.025$ (a total of 668 discretization points across both of the rings), using a timestep of $\Delta_{VF} = 1.25\times 10^{-5}$. For the normal fluid, a total of $N=256^3$ mesh points were used, with a timestep of $\Delta t_{NS} = 40\Delta t_{VF}$.

\subsection*{Helical decomposition of the flow for Hopf link simulations}

\begin{figure*}
	\centering
	\includegraphics[width=0.9\textwidth]{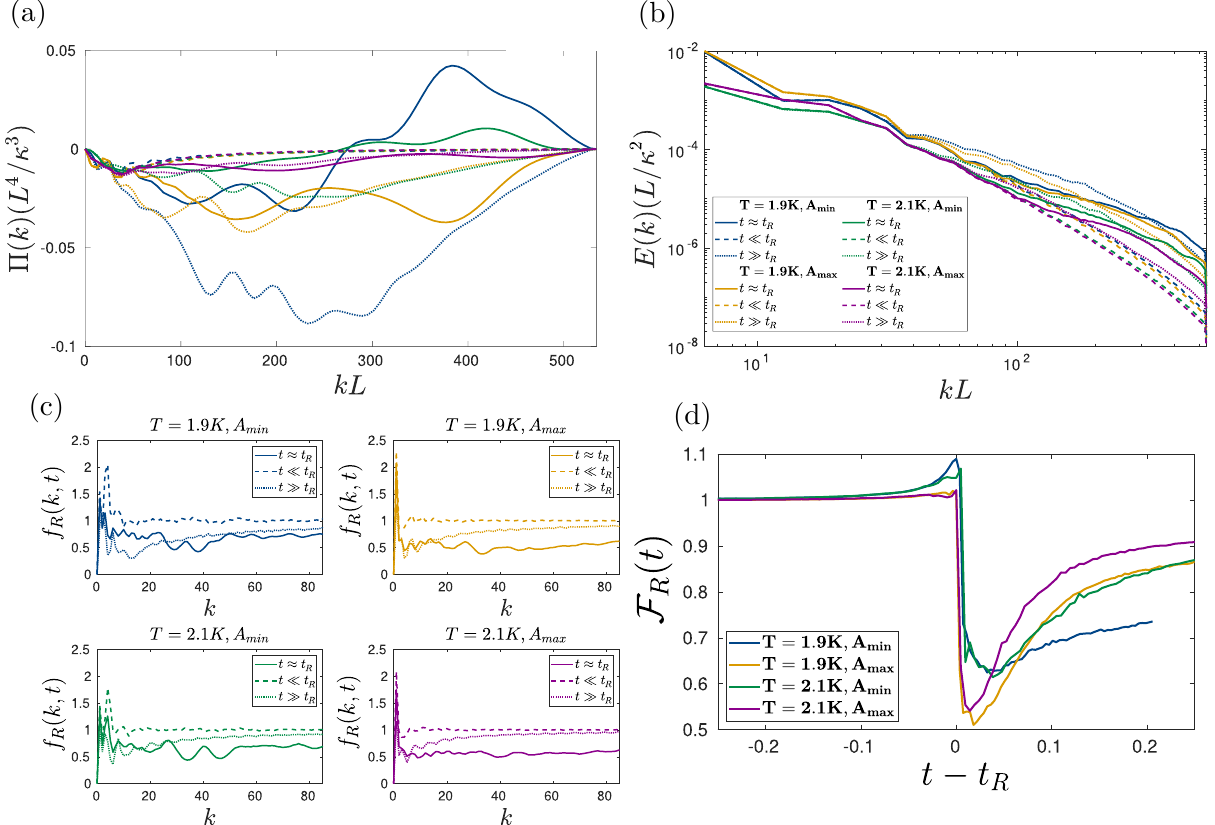}
	\caption{\emph{(a)} The scaled spectral kinetic energy flux $\Pi(k)$ for a sample of 2 Hopf link simulations across two temperatures $T=1.9$K and $T=2.1$K. The colour scheme of the lines and the linestyles is the same as for \emph{(b)}. Solid lines represent the time around reconnection, while dashed and dotted lines represent times much smaller and greater than the reconnection event respectively. \emph{(b)} The scaled kinetic energy spectrum for the same 4 Hopf link simulations. \emph{(c)} The spectrum of the ratio of the squared projected mutual friction helical modes $f_R(k,t)$, separated by temperature $T$ and the ratio of the dimensionless parameters $A^+ / A^-$. \emph{(d)} Time evolution of the squared projected helical modes $\mathcal{F}_R(t)$. }
	\label{fig:Hopf-link-decomposition}
\end{figure*}

In the main text we presented the inverse energy transfer mechanism in the context of an initially orthogonal pair of vortices, separated by an initial distance $d=0.5$ in units of the code. The reconnection of orthogonal filaments is the most violent type of vortex reconnections (the separation is much faster than the approach), and therefore shows the most clear manifestation of a chiral imbalance and hence an inverse energy transfer. However, the inverse energy transfer due to the reconnection of superfluid vortices is not limited to this geometry specifically, and will hold in general 
where the injection of helicity in the flow produces a chiral imbalance.
To illustrate this point, we perform the same analysis as for the orthogonal reconnections on a series of Hopf link reconnections, which are shown schematically in Fig.~\ref{fig:Hopf-link}. We sample two reconnections for each temperature to use for the analysis, which represent the extrema the pre/post reconnection characteristics. The minimum distance of two reconnecting filaments is well-known to scale with a $1/2$ power law \cite{stasiak2025experimental} 
\begin{equation}
	\delta^{\pm} = A^{\pm}|t - t_R|^{1/2},
\end{equation}
where the $A$ represents a dimensionless prefactor and $\pm$ is used to distinguish between the pre- and post-reconnection quantities. The ratio of $A_r = A^+/A^-$ has been shown using Gross-Pitaveskii models to be an important quantity that determines the geometric properties of vortex reconnections, such as the reconnection angle \cite{villoisIrreversibleDynamicsVortex2020}. We define $A_{min}$ and $A_{max}$ to represent two Hopf link reconnections which represent the smallest and largest differences between the approach and separation speeds of reconnecting filaments. In other words,
\begin{equation}
	A_{min} = \min \lbrace A^{+} / A^{-} \rbrace \quad A_{max} = \max \lbrace A^{+} / A^{-} \rbrace.
\end{equation}  
Therefore, $A_{min}$ is related to the reconnection event where the separation speed is similar to the approach speed and $A^+/A^- \gtrsim 1$, while $A_{max}$ is related to other extreme, where the separation speed is much greater than approach  so that $A^{+}/A^- \gg 1$. The results of the analysis is shown in Fig.~\ref{fig:Hopf-link-decomposition}, which shows a definitive chiral imbalance of mutual friction helical modes due to the reconnection event. We define the spectral ratio of the squared projected mutual friction helical modes $f_R$ by 
\begin{equation}
	f_R(k,t^*) = \frac{|f^{+}(k,t^*)|^2}{|f^{-}(k,t^*)|^2},
\end{equation}
given a fixed $t^*$, and evolution of this ratio of modes by $\mathcal{F}_R$ where
\begin{equation}
	\mathcal{F}_R(t) = \frac{\int dk |f^+(k,t)|^2}{\int dk |f^-(k,t)|^2}.
\end{equation}

\end{document}